\documentclass[10pt,aps,prd,twocolumn,showpacs,superscriptaddress,longbibliography,nofootinbib]{revtex4-2}
\usepackage{mathrsfs,amsmath,amsthm,latexsym,amssymb,amsfonts}
\usepackage{graphicx}
\usepackage{txfonts}
\usepackage{diagbox}
\usepackage[utf8]{inputenc}
\usepackage[T1]{fontenc}
\usepackage{lmodern}
\usepackage[percent]{overpic}
\usepackage{multirow}
\usepackage{booktabs}
\usepackage{xcolor}
\usepackage{appendix}
\usepackage[colorlinks=true,linkcolor=blue,citecolor=blue,urlcolor=blue]{hyperref}
\usepackage{orcidlink}

\allowdisplaybreaks
\newcommand{\ABTU}{Department of Physics, Al-Hussein Bin Talal University, 71111, Ma'an, Jordan}
\newcommand{\RGUI}{Department of Physics, The Assam Royal Global University, Guwahati-781035, Assam, India}
\newcommand{\UCCB}{Programa de P\'os-Gradua\c c\~ao em F\'{\i}sica \& Coordena\c c\~ao do Curso de F\'isica -- Bacharelado, Universidade Federal do Maranh\~{a}o, 65085-580 S\~{a}o Lu\'is, Maranh\~{a}o, Brazil}

\begin{document}
\baselineskip=12pt

\title{Hawking Temperature, Sparsity and Energy Emission Rate of Regular Black Holes Supported by Primordial Dark Matter}

\author{Faizuddin Ahmed\orcidlink{0000-0003-2196-9622}}
\email{faizuddinahmed15@gmail.com}
\affiliation{\RGUI}

\author{Ahmad Al-Badawi\orcidlink{0000-0002-3127-3453}}
\email{ahmadbadawi@ahu.edu.jo}
\affiliation{\ABTU}

\author{Edilberto O. Silva\orcidlink{0000-0002-0297-5747}}
\email{edilberto.silva@ufma.br (Corresponding author)}
\affiliation{\UCCB}

\begin{abstract}
In this paper, we investigate the thermodynamic and radiative properties of a regular black hole sourced by primordial dark matter (PDM), modeled effectively through a Dirac--Born--Infeld (DBI) scalar field. We compute the Hawking temperature, the entropy obtained from the first law at fixed PDM scale, the specific heat capacity, the sparsity parameter of the Hawking flux, and the spectral energy emission rate. Particular attention is devoted to the role played by the regularity scale parameter \(\alpha\) and to the recovery of the Schwarzschild limit. Using the normalization in which the integration constant \(M\) is the ADM mass and \(f(r)=1-2M/r+\mathcal{O}(r^{-3})\), we find that the PDM scale suppresses the Hawking temperature and the spectral energy emission rate relative to the Schwarzschild case. The fixed-\(\alpha\) heat capacity remains negative along the physical branch, indicating the persistence of local thermodynamic instability in the canonical ensemble. Moreover, within the effective-area prescription adopted here, the geometrical sparsity parameter receives a negative leading correction in the perturbative regime \(\alpha\ll 2M\), implying a slight reduction of the intermittency of the Hawking flux. We also distinguish between the near-horizon geometrical estimate and the shadow-based high-energy absorption cross-section used in the emission rate.

\end{abstract}

\maketitle

\section{Introduction}
\label{sec:1}

Regular black holes provide a useful framework for investigating possible resolutions of the curvature singularities predicted by classical general relativity. Since Bardeen's original proposal \cite{Bardeen1968}, many regular geometries have been constructed either phenomenologically or through effective matter sources, including nonlinear electrodynamics and modified-gravity models \cite{ABG1998,ABG1999,ABG2000,ABG2005,Hayward2006,Dymnikova1992,Fan2016,Balart2014,Burinskii2002,Culetu2015,Culetu2015b,Ma2015,Bronnikov2001,Lemos2011,SimpsonVisser2019,Vitalii2025,HuLanMiao2023,BronnikovFabris2006,Konoplya2026}. These solutions are especially relevant because they replace the central singularity by a finite-curvature core while preserving the main exterior features of black-hole spacetimes.

A complementary motivation comes from the possible influence of dark-matter sectors on compact objects. Recently, Parvez \cite{Parvez2026} constructed exact asymptotically flat black-hole geometries sourced by primordial dark matter distributions. In the case considered here, the matter sector is effectively modeled by a Dirac--Born--Infeld (DBI) scalar field, and the resulting spacetime is regular. The geometry contains a length scale \(\alpha\), which controls the deviation from Schwarzschild and characterizes the size of the regular core associated with the PDM source.

Thermodynamic and radiative observables offer a direct way to quantify the impact of this scale. Hawking radiation, first derived in Ref.~\cite{Hawking1975}, assigns to a black hole a temperature proportional to the surface gravity. However, the flux measured at infinity is affected by the spacetime geometry outside the horizon, and the emission is intrinsically sparse: the typical interval between successive Hawking quanta is much larger than the timescale associated with the emission of a single quantum \cite{Gray2016,Hod2016PLB,Hod2015EPJC,Miao2017PLB,Schuster2018Thesis,Visser2017MGM}. The same geometry also determines the photon sphere and the high-energy absorption cross-section, which are closely related to the black-hole shadow observed by asymptotic observers.

The goal of this paper is to investigate the thermodynamic and radiative behavior of the regular PDM-supported black hole. We derive the Hawking temperature, analyze the entropy obtained from the first law at fixed \(\alpha\), compute the specific heat capacity, study the sparsity of Hawking radiation, and evaluate the spectral energy emission rate. We also derive the photon-sphere condition and the shadow radius, emphasizing the distinction between a near-horizon geometrical estimate and the shadow-based high-energy absorption cross-section.

This paper is organized as follows. In Sec.~\ref{sec:geometry}, we review the regular PDM black-hole geometry, discuss the horizon structure, and verify the finite-curvature behavior near the center. In Sec.~\ref{sec:2}, we study the thermodynamics, including temperature, entropy, and heat capacity. In Sec.~\ref{sec:shadow}, we derive the photon-sphere condition and shadow radius. In Sec.~\ref{sec:3}, we analyze the sparsity of the Hawking flux. In Sec.~\ref{sec:4}, we discuss the spectral energy emission rate and distinguish between the near-horizon and shadow-based absorption estimates. Finally, Sec.~\ref{sec:5} contains our concluding remarks.

\section{Regular PDM black-hole geometry}
\label{sec:geometry}

The line element describing the regular PDM black hole is written as \cite{Parvez2026}
\begin{equation}
    ds^2 = -f(r)\,dt^2 + \frac{dr^2}{f(r)} + \rho^2(r)\,d\Omega^2,
    \label{metric}
\end{equation}
where
\begin{equation}
    d\Omega^2 = d\theta^2 + \sin^2\theta\,d\varphi^2,
    \qquad
    \rho(r)=\sqrt{r^2+\alpha^2}.
\end{equation}
The parameter \(\alpha\) has dimensions of length and sets the characteristic regularity scale associated with the PDM distribution, while \(M\) denotes the ADM mass. The ratio \(\alpha/M\) therefore measures the departure from the Schwarzschild geometry. In the figures below we use representative values of \(\alpha/M\) to illustrate this departure; the perturbative regime corresponds to \(\alpha/M\ll1\), whereas values of order unity probe the genuinely non-Schwarzschild domain of the effective model. Following Ref.~\cite{Parvez2026}, and setting \(G=1\), we write the lapse function as
\begin{align}
     f(r)= 1 + \frac{3 M}{\alpha}
     \left[
     \frac{r}{\alpha}
     -\left(1+\frac{r^2}{\alpha^2}\right)
     \arctan \frac{\alpha}{r}
     \right].
     \label{function}
\end{align}
Indeed, for \(r\to\infty\), one obtains
\begin{equation}
    f(r)=1-\frac{2 M}{r}+\frac{2M\alpha^2}{5r^3}
    +\mathcal{O}(r^{-5}),
    \label{function-2}
\end{equation}
so that the geometry is asymptotically flat and the Schwarzschild term is recovered at leading order. In the regime \(\alpha\ll r\), the lapse function becomes
\begin{equation}
    f(r) \approx 1-\frac{2 M}{r}+\frac{2 M \alpha^2}{5 r^3}.
    \label{function-3}
\end{equation}
This corresponds to the Schwarzschild metric plus a small \(\alpha^2\)-correction term. The perturbative dynamics, quasinormal modes, and scattering of this particular black hole have recently been studied in Refs.~\cite{Lutfuoglu2026,Skvortsova2026,Bolokhov2026}.

The event horizon \(r_h\) is determined by the condition \(f(r_h)=0\). It is convenient to introduce the dimensionless variable
\begin{equation}
    x=\frac{r_h}{\alpha},
\end{equation}
and the auxiliary function
\begin{equation}
    B(x)=
    \left(1+x^2\right)\arctan\!\frac{1}{x}
    -x.
    \label{Bdef}
\end{equation}
For \(x>0\), \(B(x)>0\). The horizon condition then gives
\begin{equation}
    M=\frac{\alpha}{3B(x)}.
    \label{aa0}
\end{equation}
For \(x\gg1\), one has \(B(x)=2/(3x)+\mathcal{O}(x^{-3})\), and Eq.~\eqref{aa0} gives \(M\simeq r_h/2\), as required by the Schwarzschild limit.

The physical branch considered in this work is therefore characterized by
\begin{equation}
    x>0,\qquad \alpha>0,\qquad M>0.
\end{equation}
For fixed \(\alpha\), Eq.~\eqref{aa0} implies a lower mass bound,
\begin{equation}
    M_{\rm min}=\lim_{x\to0^+}M(x)=\frac{2\alpha}{3\pi}.
    \label{Mmin}
\end{equation}
The coordinate horizon radius tends to zero in this limit, while the areal radius tends to \(\rho_h\to\alpha\). Thus, the geometry suggests a finite-size limiting configuration. Whether this limiting configuration should be interpreted as a true evaporation remnant requires a dynamical analysis beyond the equilibrium thermodynamic treatment adopted here.

\subsection{Near-center regularity}\label{subsec:regularity}

Although the regularity of the spacetime follows from the construction in Ref.~\cite{Parvez2026}, it is useful to make the near-center behavior explicit. For \(r/\alpha\ll1\), the lapse function behaves as
\begin{equation}
    f(r)=1-\frac{3\pi M}{2\alpha}
    +\frac{6M}{\alpha^2}r
    -\frac{3\pi M}{2\alpha^3}r^2
    +\mathcal{O}(r^3),
    \label{f_near_center}
\end{equation}
and the areal radius satisfies \(\rho(r)=\alpha+r^2/(2\alpha)+\mathcal{O}(r^4)\). Thus the two-sphere does not shrink to zero area at \(r=0\), but instead approaches the finite radius \(\alpha\).

The curvature invariants are finite in the same limit. For example, a direct expansion gives
\begin{equation}
    R\big|_{r\to0}
    =\frac{9\pi M\alpha-2\alpha^2}{\alpha^4}
    =\frac{9\pi M-2\alpha}{\alpha^3},
    \label{Ricci_center}
\end{equation}
and
\begin{equation}
    R_{\mu\nu}R^{\mu\nu}\big|_{r\to0}
    =\frac{27\pi^2M^2-18\pi M\alpha+4\alpha^2}{\alpha^6}.
    \label{Ricci_sq_center}
\end{equation}
The Kretschmann scalar is also finite. Hence the central region is characterized by finite curvature controlled by the scale \(\alpha\), confirming that the geometry is regular on the branch considered here.

Within the physical branch, the thermodynamic quantities can be expressed as functions of \(x\) and \(\alpha\). This parametrization is useful because it separates the overall length scale from the dimensionless dependence on the regularity parameter.

\begin{figure}[t]
    \centering
    \includegraphics[width=\columnwidth]{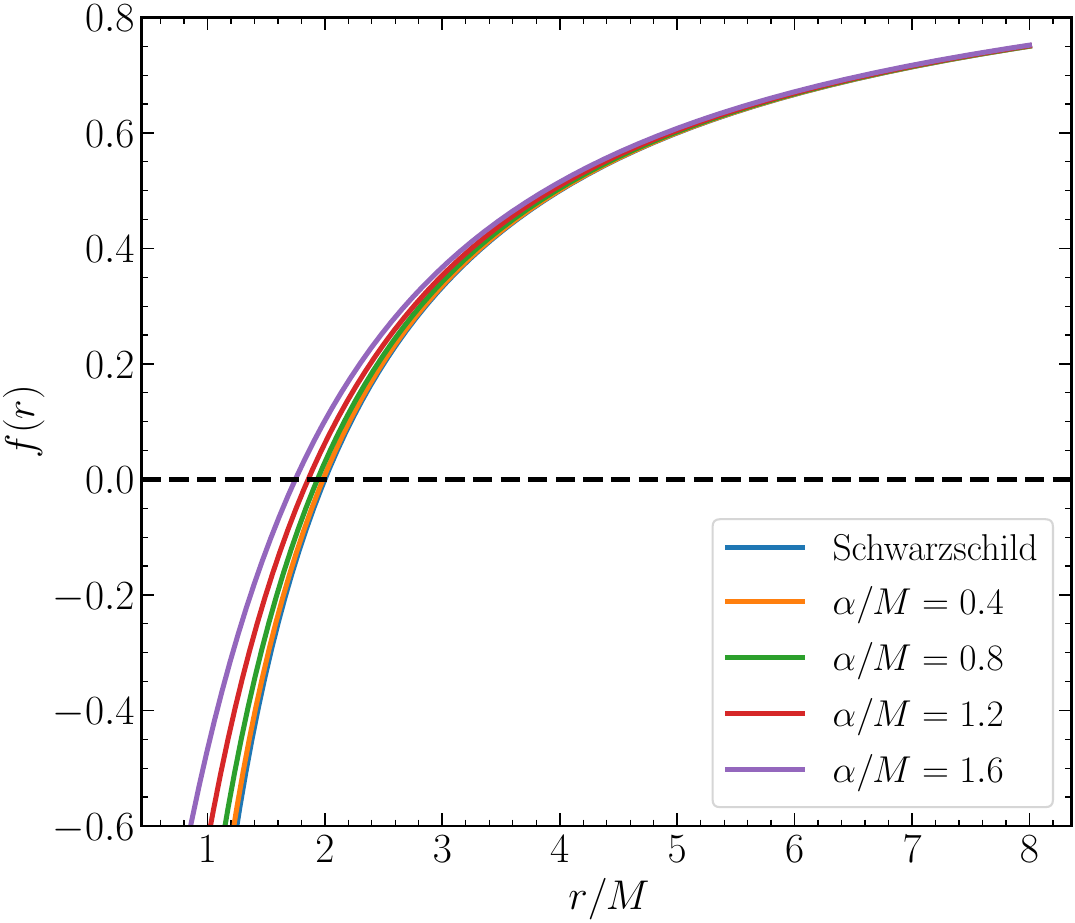}
    \caption{Metric function \(f(r)\) as a function of \(r/M\) for selected values of \(\alpha/M\). The horizontal dashed line indicates \(f(r)=0\), whose intersections with the curves determine the horizon radius. The Schwarzschild result is shown for comparison.}
    \label{fig:metric_function}
\end{figure}

Figure~\ref{fig:metric_function} shows the behavior of the lapse function as a function of the dimensionless coordinate \(r/M\) for different values of \(\alpha/M\). The Schwarzschild curve is recovered when \(\alpha=0\). For nonzero \(\alpha\), the regularity scale shifts the horizon location while preserving the asymptotically flat behavior at large distances.

\begin{figure}[t]
\centering
\includegraphics[width=\columnwidth]{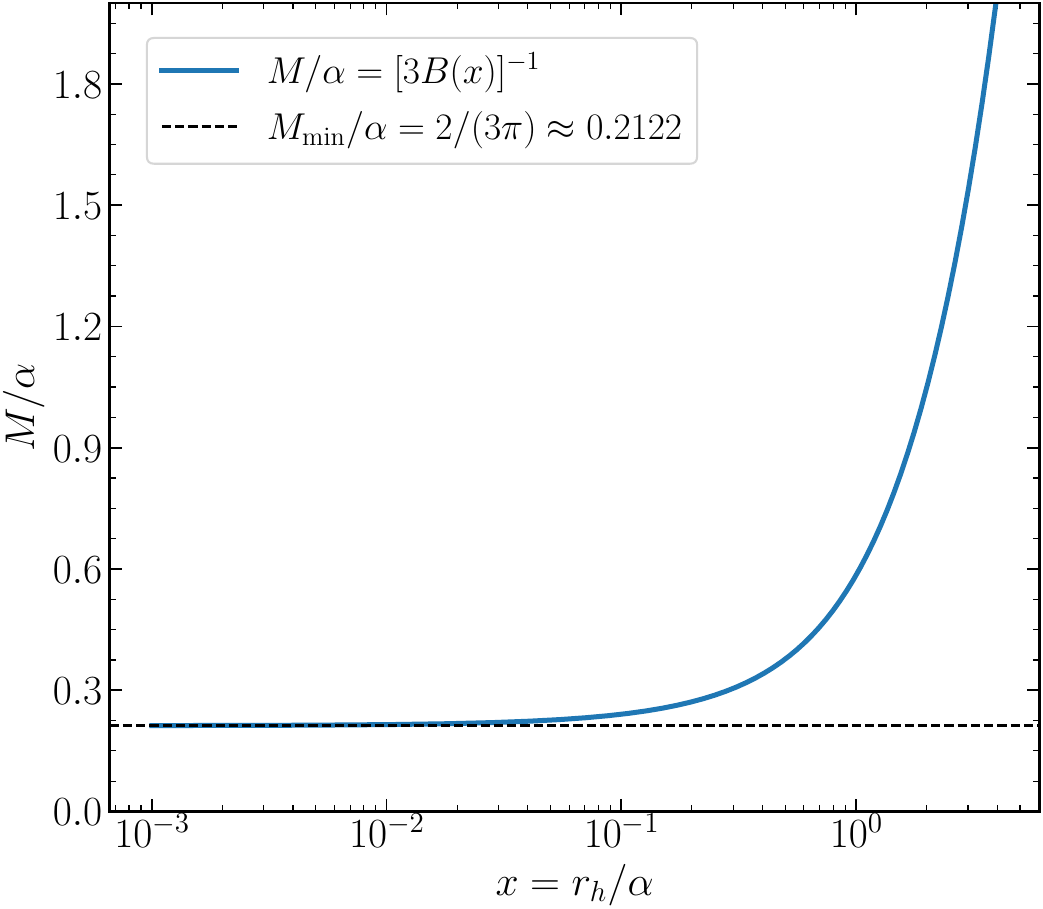}
    \caption{Dimensionless ADM mass \(M/\alpha\) as a function of \(x=r_h/\alpha\). The dashed horizontal line marks the lower bound \(M_{\rm min}/\alpha=2/(3\pi)\).}
    \label{fig:mass_vs_x}
\end{figure}

The mass function is displayed in Fig.~\ref{fig:mass_vs_x}. The curve approaches the Schwarzschild relation \(M/\alpha\simeq x/2\) for \(x\gg1\), whereas for \(x\to0^+\) it tends to the finite value \(M_{\rm min}/\alpha=2/(3\pi)\). This illustrates the lower mass bound associated with the regular PDM-supported configuration.

\section{Thermodynamics}
\label{sec:2}

We now investigate the thermodynamics of the PDM-supported regular black hole. Throughout this section, derivatives with respect to \(r_h\) are taken at fixed \(\alpha\), unless otherwise stated.

\begin{figure*}[tbhp]
    \centering
    \includegraphics[scale=0.6]{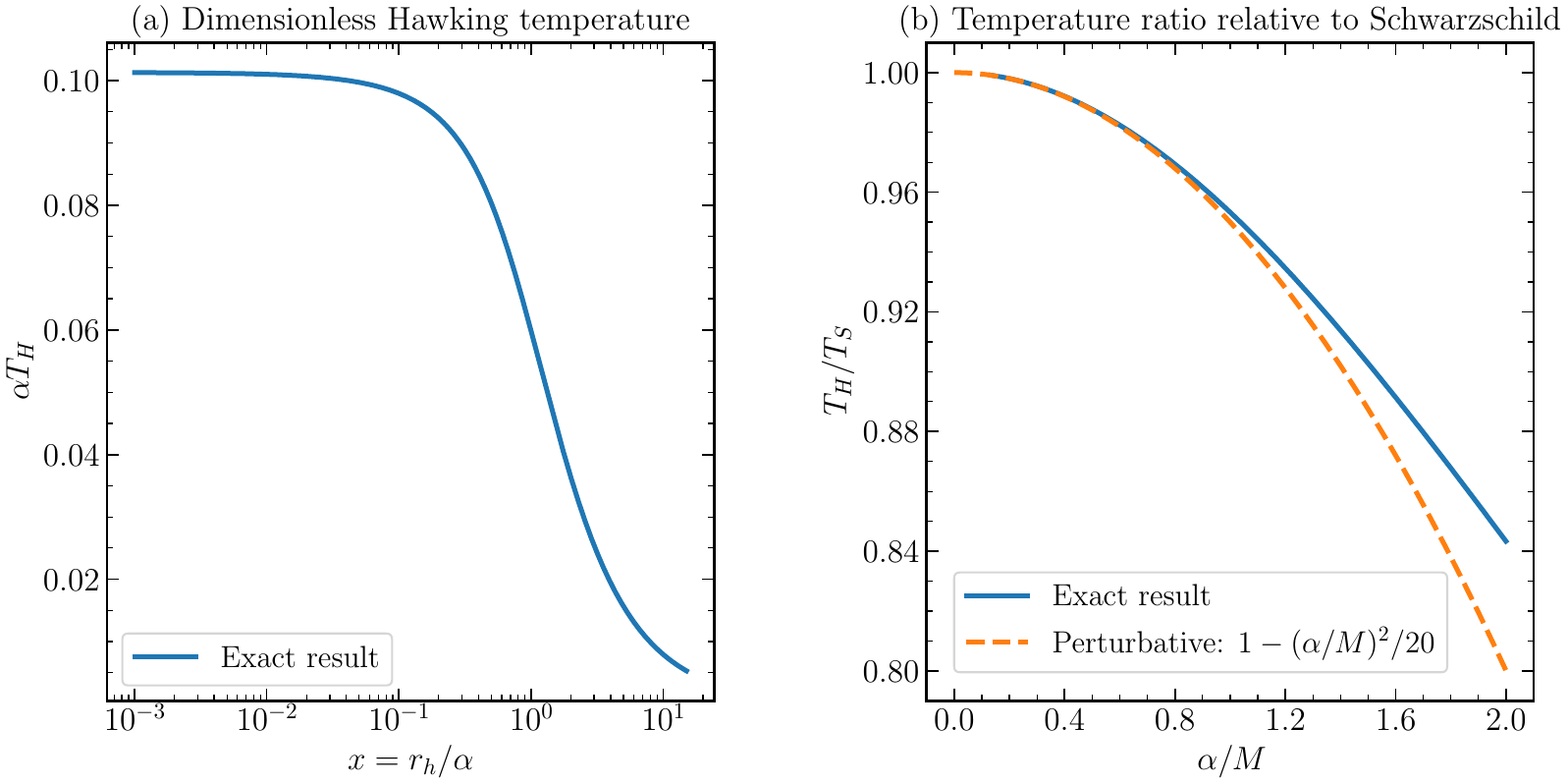}
    \caption{Hawking temperature. Panel (a) shows the exact dimensionless temperature \(\alpha T_H\) as a function of \(x=r_h/\alpha\). Panel (b) compares the exact ratio \(T_H/T_S\) with the perturbative expression \(1-(\alpha/M)^2/20\). The PDM scale lowers the temperature relative to the Schwarzschild value at fixed ADM mass.}
    \label{fig:temperature}
\end{figure*}

\subsection{Hawking temperature}
\label{subsec:temperature}

The surface gravity is defined by \cite{Hawking1975,BardeenCarterHawking1973,Wald1994}
\begin{equation}
\kappa = -\frac{1}{2}\lim_{r \to r_h}\frac{\partial_r g_{tt}}{\sqrt{-g_{tt}\,g_{rr}}}=\frac{1}{2}\frac{\partial f}{\partial r}\bigg|_{r=r_h}.
\label{aa1}
\end{equation}
Therefore, the Hawking temperature \(T_H=\kappa/(2\pi)\) is
\begin{equation}
T_H=\frac{1}{2\pi\alpha}\,\frac{1-x\arctan(1/x)}{\left(1+x^2\right)\arctan(1/x)-x}.\label{aa2}
\end{equation}
Equivalently, in terms of \(r_h\) and \(\alpha\),
\begin{equation}
T_H=\frac{1}{2\pi\alpha}\frac{1-\frac{r_h}{\alpha}\arctan\!\frac{\alpha}{r_h}}{\left(1+\frac{r_h^2}{\alpha^2}\right)\arctan\!\frac{\alpha}{r_h}-\frac{r_h}{\alpha}}.
\end{equation}
Since \(1-x\arctan(1/x)>0\) and \(B(x)>0\) for \(x>0\), the Hawking temperature is positive along the physical branch.

The Schwarzschild limit is recovered by expanding for \(x\gg1\), or equivalently \(\alpha/r_h\ll1\). In this regime,
\begin{equation}
T_H=\frac{1}{4\pi r_h}\left(1-\frac{2\alpha^2}{5r_h^2}+\mathcal{O}\!\left(\frac{\alpha^4}{r_h^4}\right)\right).
\end{equation}
Using the perturbative relation between \(r_h\) and \(M\), obtained below in Eq.~\eqref{ff1}, this becomes
\begin{equation}
T_H \approx T_S \left(1-\frac{\alpha^2}{20M^2}\right), \qquad T_S=\frac{1}{8\pi M}.\label{ff2}
\end{equation}
Thus, for fixed ADM mass, the PDM regularity scale suppresses the Hawking temperature relative to the Schwarzschild value.

This behavior is summarized in Fig.~\ref{fig:temperature}. Panel~(a) shows the exact dimensionless quantity \(\alpha T_H\) as a function of \(x=r_h/\alpha\). The curve remains finite in the small-\(x\) limit and approaches the Schwarzschild scaling at large \(x\). Panel~(b) displays the ratio \(T_H/T_S\) as a function of \(\alpha/M\), comparing the exact result with the perturbative approximation in Eq.~\eqref{ff2}. The agreement at small \(\alpha/M\) confirms the consistency of the expansion, while the exact curve captures the deviation outside the strictly perturbative regime.

\subsection{Entropy from the first law}\label{subsec:entropy}

The area of the horizon is
\begin{equation}
\mathcal{A}=\lim_{r \to r_h}\int \int \sqrt{g_{\theta\theta}\,g_{\phi\phi}}\,d\theta d\phi=4 \pi (r_h^2+\alpha^2).\label{aa3}
\end{equation}
Because the spacetime contains an effective PDM/DBI source and the areal radius is \(\rho(r)=\sqrt{r^2+\alpha^2}\), it is important not to assume from the outset that the thermodynamic entropy is simply given by \(\mathcal{A}/4\). Instead, we define the entropy through the first law at fixed \(\alpha\),
\begin{equation}
    dM=T_H\,dS.
\end{equation}
Using Eq.~\eqref{aa0}, one obtains
\begin{equation}
\frac{dS}{dr_h}=\frac{1}{T_H}\left(\frac{\partial M}{\partial r_h}\right)_\alpha=\frac{4\pi \alpha}{3B(x)}.
\end{equation}
Therefore,
\begin{equation}
S(r_h,\alpha)=\frac{4\pi\alpha}{3}\int\frac{dr_h}{\left(1+\frac{r_h^2}{\alpha^2}\right)\arctan\!\frac{\alpha}{r_h}-\frac{r_h}{\alpha}}+S_0,\label{entropy_int}
\end{equation}
or, equivalently,
\begin{equation}
S(x,\alpha)=\frac{4\pi\alpha^2}{3}\int^x\frac{d\bar x}{B(\bar x)}+S_0.\label{entropy_int_x}
\end{equation}
where the integration is performed at fixed \(\alpha\), and \(S_0\) is an integration constant. A natural choice is to fix \(S_0\) so that the entropy reproduces the Schwarzschild behavior in the large-\(x\) regime.

For \(x\gg1\), Eq.~\eqref{entropy_int} gives
\begin{equation}
\frac{dS}{dr_h}=2\pi r_h\left[1+\frac{\alpha^2}{5r_h^2}+\mathcal{O}\!\left(\frac{\alpha^4}{r_h^4}\right)\right],
\end{equation}
and hence
\begin{equation}
S(r_h,\alpha)=\pi r_h^2+\frac{2\pi\alpha^2}{5}\ln\!\left(\frac{r_h}{\ell_0}\right)+\cdots,\label{entropy_expansion}
\end{equation}
where \(\ell_0\) is a constant length scale associated with the integration constant. The leading term is the usual Schwarzschild entropy, while the subleading terms encode the effect of the regular PDM scale. This shows explicitly that the entropy obtained from the first law is not identical to \(\mathcal{A}/4=\pi(r_h^2+\alpha^2)\), except in the leading Schwarzschild limit.

\subsection{Specific heat and local stability}
\label{subsec:specific_heat}

The fixed-\(\alpha\) heat capacity is defined as
\begin{equation}
C_\alpha=\left(\frac{\partial M}{\partial T_H}\right)_\alpha=\frac{\left(\partial M/\partial x\right)_\alpha}{\left(\partial T_H/\partial x\right)_\alpha}.
\end{equation}
Using Eqs.~\eqref{aa0} and \eqref{aa2}, we find
\begin{equation}
C_\alpha=\frac{4\pi\alpha^2}{3}\,\frac{U(x)}{Q(x)},\label{specific_heat_corrected}
\end{equation}
where
\begin{equation}
U(x)=1-x\arctan\!\frac{1}{x},
\end{equation}
and
\begin{align}
Q(x)=\left[\frac{x}{1+x^2}-\arctan\!\frac{1}{x}\right]B(x)+2U^2(x).\label{Qdef}
\end{align}
For \(x>0\), \(U(x)>0\). A direct sign analysis of Eq.~\eqref{Qdef} shows that \(Q(x)<0\) along the physical branch. Therefore,
\begin{equation}
C_\alpha<0\qquad(x>0).
\end{equation}
The regular PDM scale modifies the magnitude of the heat capacity but does not yield a locally stable canonical branch for the metric normalization adopted here. In particular, no Davies-type divergence occurs in the physical domain because \(Q(x)\) does not vanish for \(x>0\).

In the Schwarzschild regime \(x\gg1\), one obtains
\begin{equation}
C_\alpha=-2\pi r_h^2\left[1+\mathcal{O}\!\left(\frac{\alpha^2}{r_h^2}\right)\right],
\end{equation}
which reproduces the usual negative heat capacity of the Schwarzschild black hole at leading order.

\begin{figure*}[ht!]
    \centering
    \includegraphics[scale=0.65]{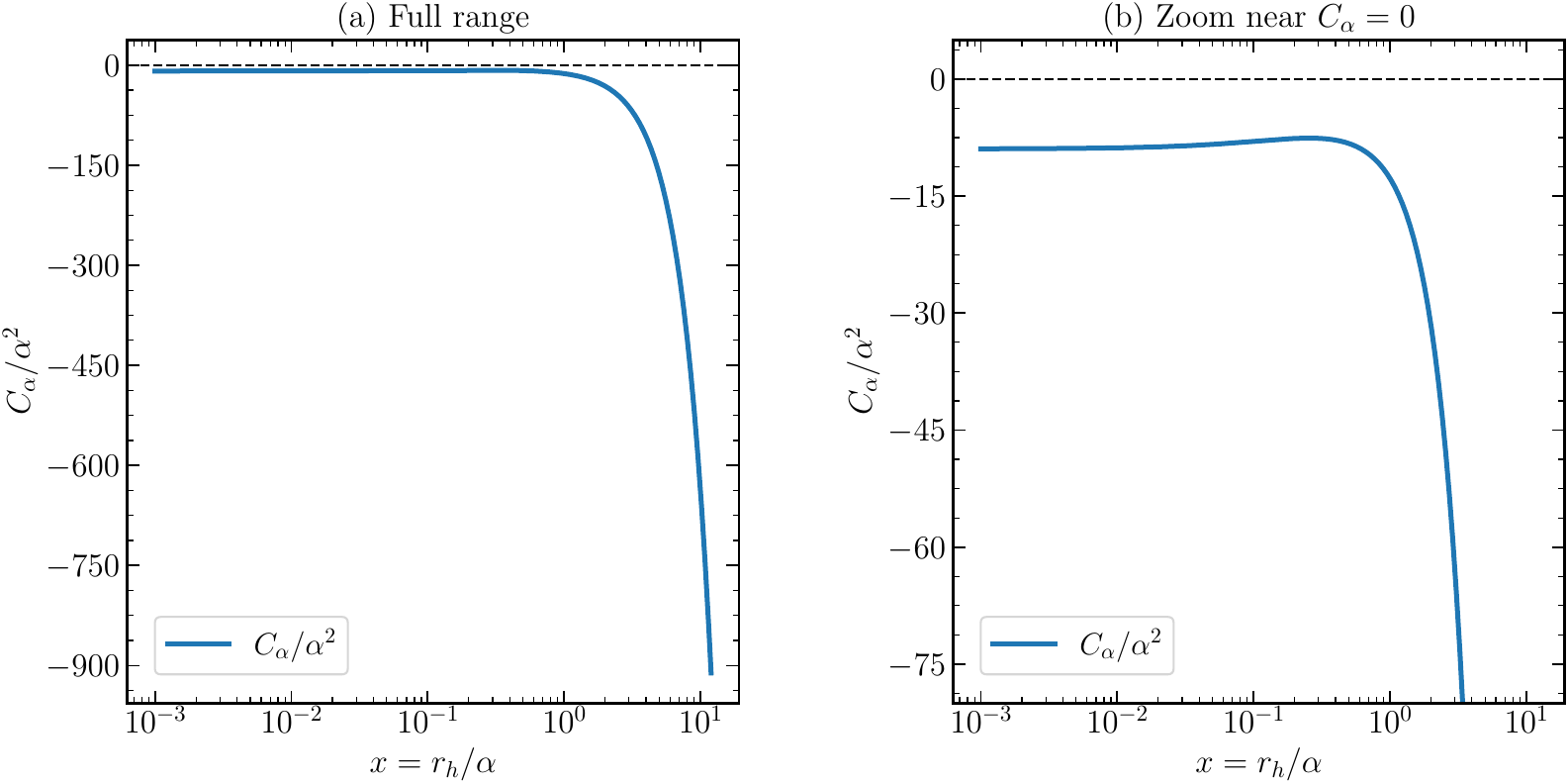}
    \caption{Fixed-\(\alpha\) heat capacity \(C_\alpha/\alpha^2\) as a function of \(x=r_h/\alpha\). The curve remains negative along the physical branch, indicating local thermodynamic instability in the canonical ensemble.}
    \label{fig:heat_capacity}
\end{figure*}

The sign of the heat capacity is illustrated in Fig.~\ref{fig:heat_capacity}. The plotted quantity \(C_\alpha/\alpha^2\) remains negative throughout the physical branch, supporting the analytical conclusion that the canonical instability persists in this model. The absence of a pole in the plotted domain also confirms that no Davies-type phase transition occurs for \(x>0\) under the fixed-\(\alpha\) prescription.

\subsection{Comment on extended thermodynamics}\label{subsec:extended}

If the scale parameter \(\alpha\) is allowed to vary, the mass may be regarded as a function \(M=M(S,\alpha)\), and one may formally write an extended first law,
\begin{equation}
    dM=T\,dS+\Phi\,d\alpha.\label{ww2}
\end{equation}
At fixed \(\alpha\), the thermodynamic temperature is
\begin{equation}
T=\left(\frac{\partial M}{\partial S}\right)_\alpha=T_H.\label{ww3}
\end{equation}
However, the potential entering Eq.~\eqref{ww2} must be defined at fixed entropy,
\begin{equation}
    \Phi=\left(\frac{\partial M}{\partial\alpha}\right)_S.
\end{equation}
This is not the same as the derivative at fixed horizon coordinate \(r_h\). For reference, the latter quantity is
\begin{equation}
\Phi_{r_h}=\left(\frac{\partial M}{\partial \alpha}\right)_{r_h}=\frac{1}{3B(x)}+\frac{x}{3B^2(x)}\frac{dB(x)}{dx},\label{ww4}
\end{equation}
with
\begin{equation}
\frac{dB}{dx}=2x\arctan\!\frac{1}{x}-2.
\end{equation}
Equation~\eqref{ww4} should therefore be interpreted only as the fixed-\(r_h\) response of the mass to a variation in \(\alpha\). The thermodynamic conjugate \(\Phi=(\partial M/\partial\alpha)_S\) requires the explicit entropy function \(S(r_h,\alpha)\), including the prescription for the integration constant.

The same caution applies to the Smarr relation. Although \(M\) is homogeneous of degree one under the simultaneous scaling \((r_h,\alpha)\to(\lambda r_h,\lambda\alpha)\), a formal expression of the type
\begin{equation}
M=2TS+\alpha\Phi\label{ww5}
\end{equation}
cannot be imposed as an independent identity without specifying \(S(r_h,\alpha)\) and the corresponding fixed-entropy potential. In the present work, we therefore regard the fixed-\(\alpha\) first law as the primary thermodynamic relation and treat the extended relation as a formal observation.
\begin{figure*}[tbhp]
    \centering
    \includegraphics[scale=0.6]{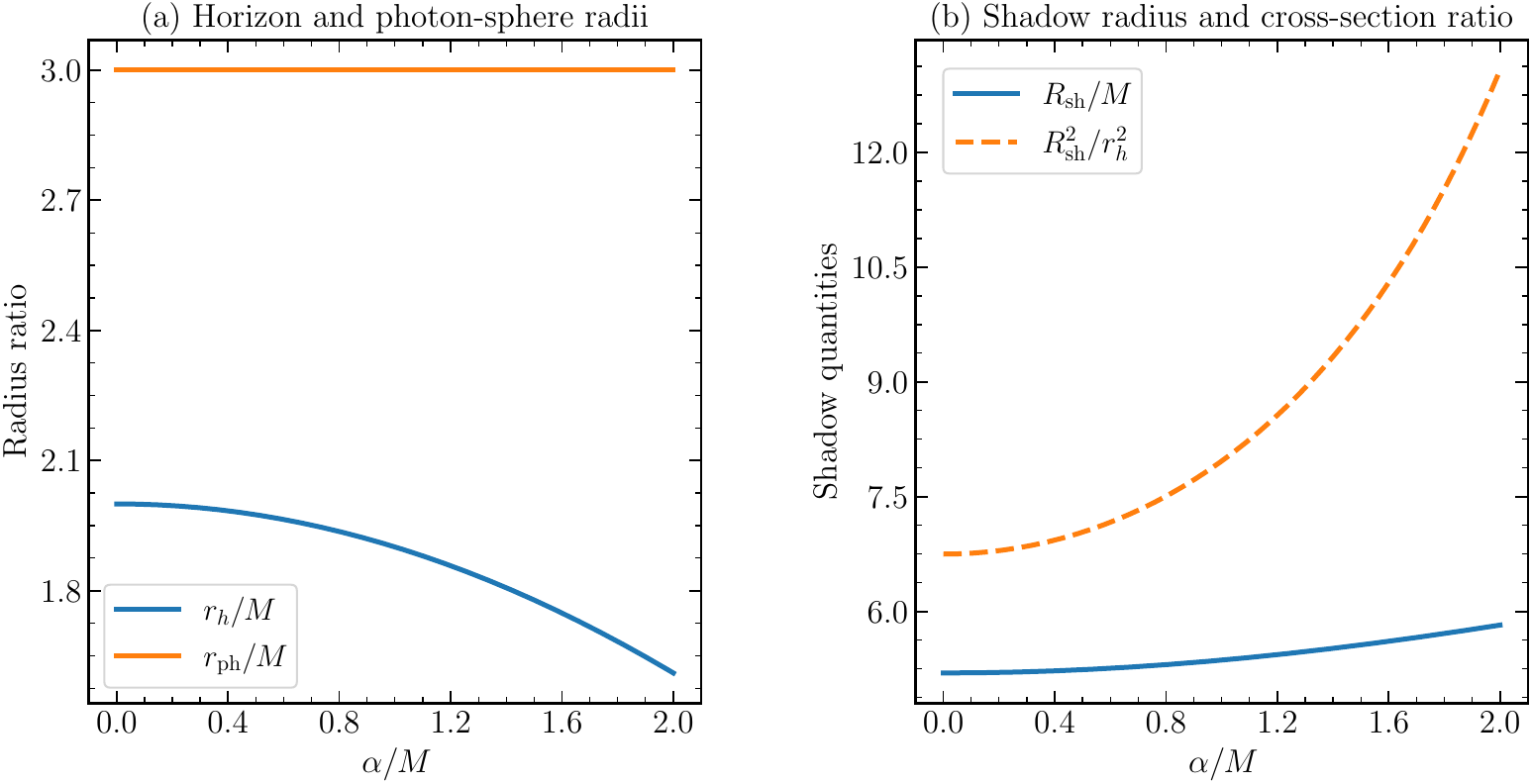}
    \caption{Shadow-related quantities. Panel (a) shows \(r_h/M\) and \(r_{\rm ph}/M\) as functions of \(\alpha/M\). Panel (b) displays the shadow radius \(R_{\rm sh}/M\) and the cross-section ratio \(R_{\rm sh}^2/r_h^2\). These quantities determine the difference between the horizon-area and shadow-based estimates of the absorption cross-section.}
    \label{fig:shadow_quantities}
\end{figure*}

\section{Photon sphere and black-hole shadow}\label{sec:shadow}

For completeness and to ensure the normalization entering the high-energy emission rate is correct, we summarize the null circular-orbit construction for the metric \eqref{metric}. Since the areal radius is
\begin{equation}
    \rho(r)=\sqrt{r^2+\alpha^2},
\end{equation}
the optical radius relevant for photon motion is not simply \(r\), but rather \(\rho(r)/\sqrt{f(r)}\).

Restricting the motion to the equatorial plane, the conserved energy and angular momentum of a photon are
\begin{equation}
    E=f(r)\dot t,
    \qquad
    L=\rho^2(r)\dot\varphi,
\end{equation}
where the dot denotes differentiation with respect to an affine parameter. The null condition gives
\begin{equation}
    \dot r^{\,2}=E^2-\frac{f(r)L^2}{\rho^2(r)}
    \equiv E^2-V_{\rm eff}(r).
\end{equation}
The radius \(r_{\rm ph}\) of an unstable circular photon orbit is determined by
\begin{equation}
    V_{\rm eff}(r_{\rm ph})=E^2,
    \qquad
    V'_{\rm eff}(r_{\rm ph})=0.
\end{equation}
Equivalently,
\begin{equation}
    \left.
    \frac{d}{dr}
    \left(\frac{f(r)}{\rho^2(r)}\right)
    \right|_{r=r_{\rm ph}}
    =0,
\end{equation}
or
\begin{equation}
    \frac{f'(r_{\rm ph})}{f(r_{\rm ph})}
    =
    \frac{2r_{\rm ph}}{r_{\rm ph}^2+\alpha^2}.
    \label{photon_condition}
\end{equation}
The corresponding critical impact parameter, which fixes the shadow radius as measured by an observer at infinity, is
\begin{equation}
    b_c^2=R_{\rm sh}^2
    =
    \frac{\rho^2(r_{\rm ph})}{f(r_{\rm ph})}
    =
    \frac{r_{\rm ph}^2+\alpha^2}{f(r_{\rm ph})}.
    \label{shadow_radius}
\end{equation}
Equations~\eqref{photon_condition} and \eqref{shadow_radius} are the appropriate inputs for the high-energy absorption cross-section,
\begin{equation}
    \sigma_{\rm lim}\simeq \pi R_{\rm sh}^2.
\end{equation}
In the Schwarzschild limit, \(r_{\rm ph}=3M\) and \(R_{\rm sh}=3\sqrt{3}M\), so that \(\sigma_{\rm lim}=27\pi M^2\), as expected.

\begin{table}[h]
\centering
\begin{tabular}{|c|c|c|c|c|c|}
\hline
$\alpha$ & $0.4$ & $0.8$ & $1.2$ & $1.6$ & $2.0$ \\ \hline
$R_{sh}$ & 5.22374 & 5.30499 & 5.43582 & 5.61041 & 5.82232 \\ \hline
\end{tabular}
\caption{The shadow radius has been tabulated numerically for a regular black hole with PDM.}
\label{table2a}
\end{table}

The numerical behavior of the horizon, photon sphere, and shadow quantities is shown in Fig.~\ref{fig:shadow_quantities}. Panel~(a) compares the horizon radius with the photon-sphere radius. In the present normalization, the photon-sphere coordinate radius remains equal to its Schwarzschild value \(r_{\rm ph}=3M\), while the horizon radius decreases as \(\alpha/M\) increases. This does not mean that the observable shadow is unchanged: because the optical radius is \(\rho(r)/\sqrt{f(r)}\), the shadow radius depends on \(\alpha\) even when the coordinate location of the photon sphere does not. Panel~(b) shows the shadow radius and the ratio \(R_{\rm sh}^2/r_h^2\), which measures how much the shadow-based high-energy cross-section differs from the simple horizon-area estimate.

\begin{figure*}[tbhp]
    \centering
    \includegraphics[scale=0.6]{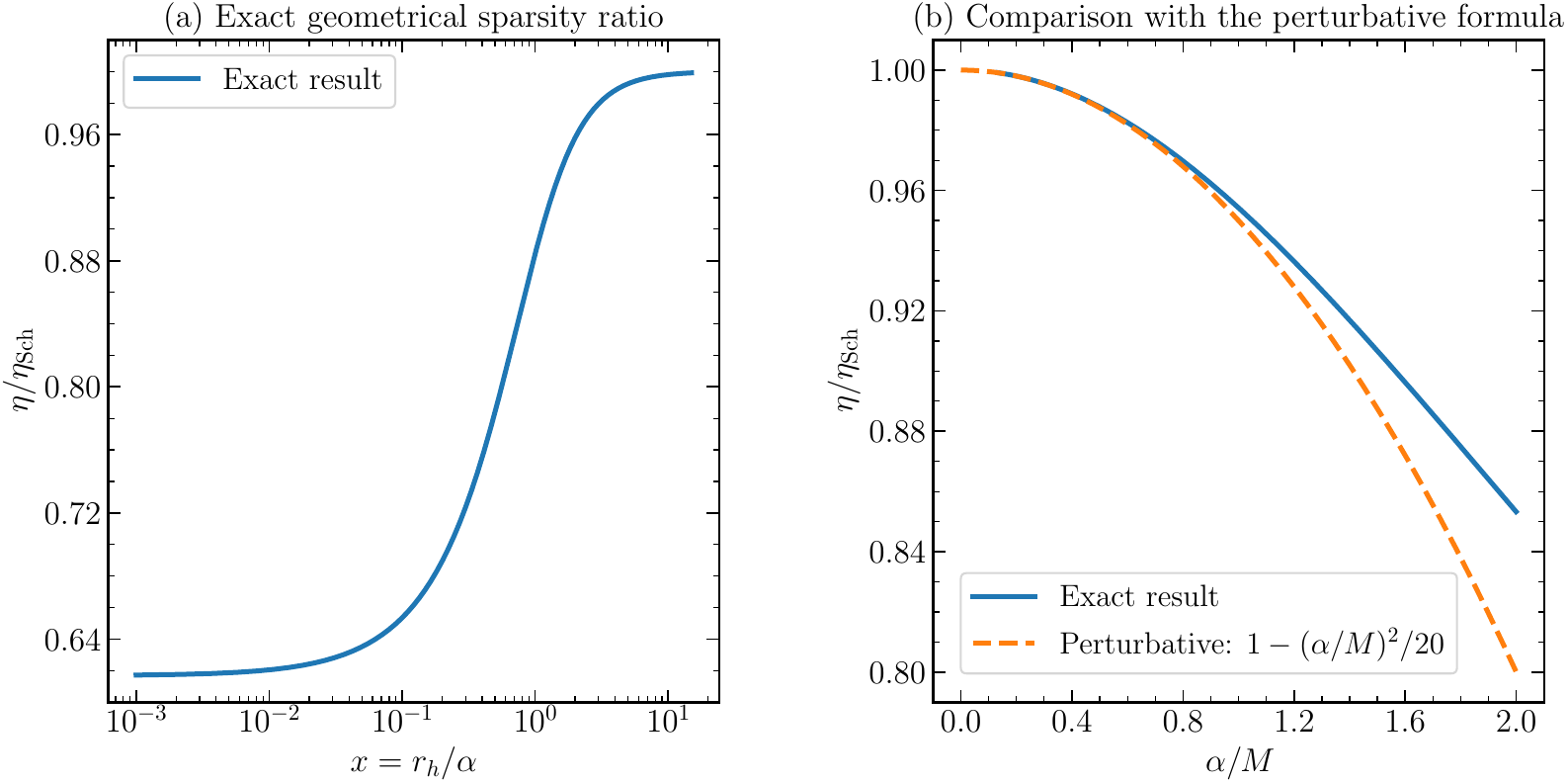}
    \caption{Geometrical sparsity ratio. Panel (a) shows the exact ratio \(\eta/\eta_{\rm Sch}\) as a function of \(x=r_h/\alpha\). Panel (b) compares the exact result with the perturbative correction \(1-(\alpha/M)^2/20\). Within the adopted effective-area prescription, the PDM scale slightly reduces sparsity compared with the Schwarzschild value.}
    \label{fig:sparsity}
\end{figure*}

\section{Sparsity of radiation}\label{sec:3}

Although Hawking radiation exhibits a thermal spectrum, the emission process is not continuous in time. Instead, it proceeds through the emission of discrete, well-separated quanta, implying that the Hawking flux is intrinsically sparse \cite{Visser2017MGM,Gray2016}. A useful way to quantify this sparsity is to compare the typical thermal wavelength of the emitted particles with the black hole's effective emission area \cite{Visser2017MGM}.

The sparsity parameter is defined as
\begin{equation}
    \eta=
    \frac{\mathcal{C}}{\tilde{g}}\,
    \frac{\lambda_{\rm th}^2}{A_{\rm eff}},
    \label{bb1}
\end{equation}
where \(\mathcal{C}\) is a positive numerical constant, \(\tilde{g}\) is the number of spin states of the emitted quanta, and
\begin{equation}
    \lambda_{\rm th}=\frac{2\pi}{T_H},
    \qquad
    A_{\rm eff}=\frac{27}{4}\,A_{\rm BH}(r_h).
    \label{bb2}
\end{equation}
In what follows, we set \(\mathcal{C}/\tilde g=1\), so that the expressions below measure the geometrical sparsity. Other choices simply rescale \(\eta\) by a constant factor and do not change the dependence on \(r_h\) or \(\alpha\).

Using \(A_{\rm BH}=4\pi(r_h^2+\alpha^2)\) and Eq.~\eqref{aa2}, we obtain
\begin{equation}
    \eta
    =
    \frac{16\pi^3}{27}
    \frac{B^2(x)}
    {(1+x^2)U^2(x)},
    \label{bb3}
\end{equation}
where \(B(x)\) and \(U(x)\) were defined above. The corresponding Schwarzschild value is
\begin{equation}
    \eta_{\rm Sch}=\frac{64\pi^3}{27}.
    \label{ff5}
\end{equation}
For \(x\gg1\), Eq.~\eqref{bb3} gives
\begin{equation}
    \eta\to \eta_{\rm Sch}.
\end{equation}

It is useful to express the leading correction at fixed ADM mass. In the perturbative regime \(\alpha\ll2M\), the horizon radius is
\begin{equation}
    r_h \approx 2M\left(1-\frac{\alpha^2}{20 M^2}\right).
    \label{ff1}
\end{equation}
The Hawking temperature is given by Eq.~\eqref{ff2}, while the horizon area becomes
\begin{equation}
    \mathcal{A}
    \approx
    A_S
    \left(1+\frac{3 \alpha^2}{20 M^2}\right),
    \qquad
    A_S=16\pi M^2.
    \label{ff3}
\end{equation}
Consequently, within the effective-area prescription of Eq.~\eqref{bb2},
\begin{equation}
    \eta
    =
    \left(1-\frac{\alpha^2}{20 M^2}\right)
    \eta_{\rm Sch}
    +
    \mathcal{O}\!\left(\frac{\alpha^4}{M^4}\right).
    \label{ff4}
\end{equation}
Thus, in the perturbative regime, the PDM scale slightly decreases the geometrical sparsity. This occurs because the increase in the effective area overcompensates the decrease in the temperature when the combination \(A_{\rm eff}T_H^2\) is expanded to order \(\alpha^2\).

Figure~\ref{fig:sparsity} displays the exact sparsity ratio and its perturbative approximation. Panel~(a) shows \(\eta/\eta_{\rm Sch}\) as a function of \(x=r_h/\alpha\). The Schwarzschild value is approached at large \(x\), whereas smaller \(x\) values lead to reduced geometrical sparsity in the prescription used here. Panel~(b) confirms that the perturbative expression in Eq.~\eqref{ff4} accurately describes the small-\(\alpha/M\) regime.

\begin{figure*}[tbhp]
    \centering
    \includegraphics[scale=0.6]{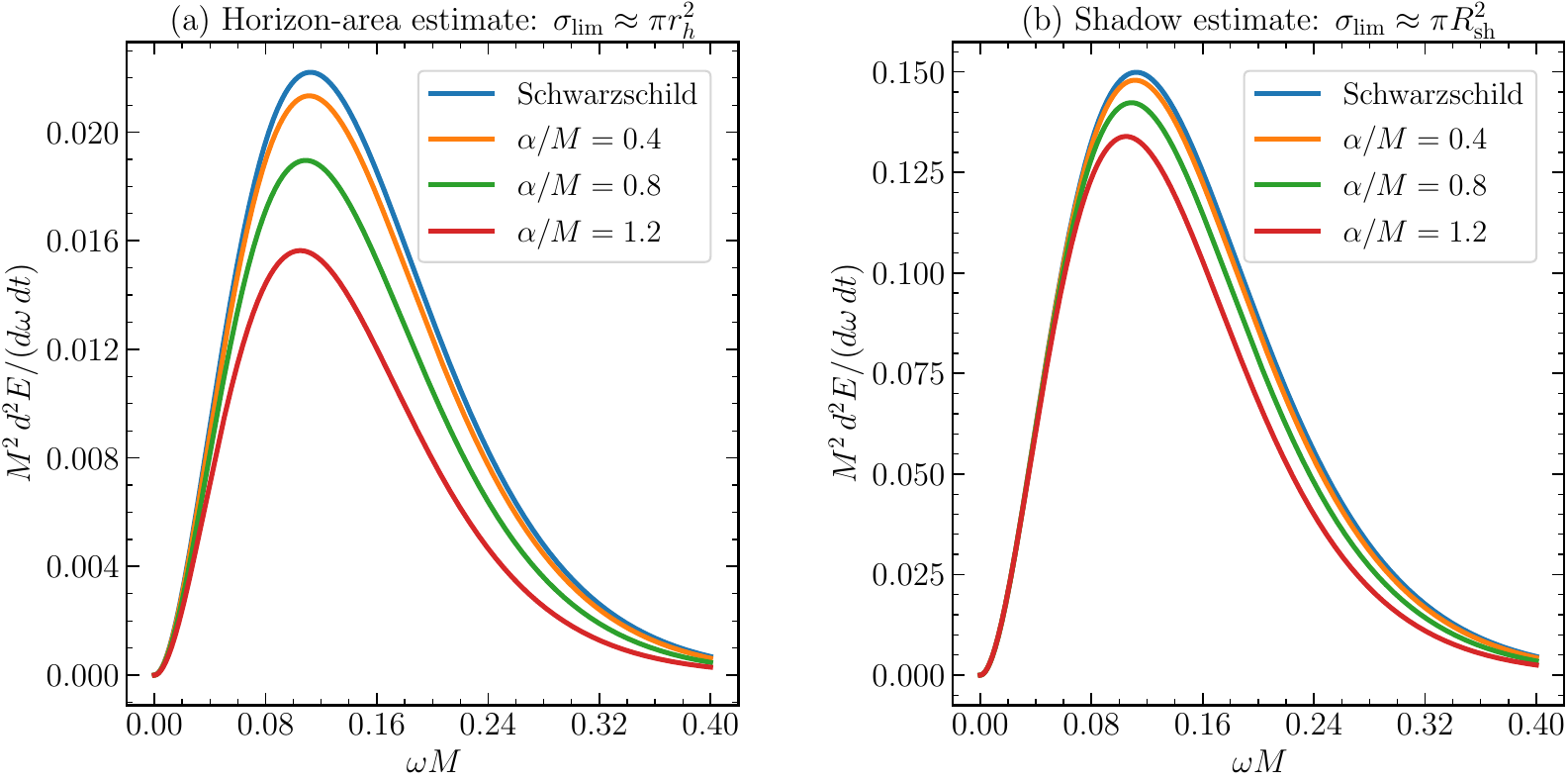}
    \caption{Spectral energy emission rate as a function of \(\omega M\) for selected values of \(\alpha/M\). Panel (a) uses the near-horizon estimate \(\sigma_{\rm lim}\approx\pi r_h^2\). Panel (b) uses the shadow-based high-energy estimate \(\sigma_{\rm lim}\approx\pi R_{\rm sh}^2\). The PDM regularity scale suppresses the emission rate relative to the Schwarzschild case.}
    \label{fig:emission_rate}
\end{figure*}

\section{Energy emission rate}\label{sec:4}

The continuous creation and annihilation of particle pairs in the vicinity of the event horizon leads to what is commonly associated with black hole radiation. In this picture, quantum fluctuations near the horizon generate particle-antiparticle pairs, with one particle tunneling out of the black hole while the other falls inward. This process results in the emission of radiation, as originally described in the context of Hawking’s semiclassical analysis \cite{Hawking1975}. The escaping positive-energy particles reduce the black hole's mass over time, leading to a gradual loss of energy and eventual evaporation. This phenomenon, known as Hawking radiation, arises from quantum effects in curved spacetime near the event horizon.

The spectral energy emission rate is usually written in terms of a limiting absorption cross-section as \cite{Wei2013,Sanchez1978,Decanini2011}
\begin{equation}
\frac{d^2\mathbb{E}}{d\omega dt} = \frac{2\pi^2\,\sigma_{\rm lim}}{e^{\omega/T_H}-1}\,\omega^3,\label{cc1}
\end{equation}
where \(\omega\) denotes the emitted frequency.

The rate of this evaporation is directly related to the rate of energy emission, which can be observed by a far-off observer as the cross-section of high-energy reception approximating the shadow/horizon of a black hole. At very high energies, the absorption cross-section oscillates around a limiting constant value ($\sigma_{\rm lim}$). This limiting value is determined by the event horizon radius \( r_h \), which sets the characteristic geometric scale of the black hole and is given by \cite{Ditta2022,Ditta2023,Javed2023,Channuie2025,Bouzenada2025}
\begin{equation}
    \sigma_{\rm lim}\approx\pi r_h^2.\label{cc2}
\end{equation}
This approximation is useful for a simple geometric estimate of the horizon scale. Using Eqs.~\eqref{ff1} and \eqref{ff2}, one obtains, to order \(\alpha^2/M^2\),
\begin{equation}
\frac{d^2\mathbb{E}}{d\omega dt}=\frac{8 \pi^3 M^2 \omega^3 \left(1-\frac{\alpha^2}{10 M^2}\right)} {\exp\left[8 \pi M \omega
\left(1+\frac{\alpha^2}{20 M^2}\right)\right]-1}.\label{cc3}
\end{equation}
The correction in the numerator comes from \(r_h^2\simeq4M^2(1-\alpha^2/10M^2)\), whereas the correction in the exponential comes from
\[\frac{1}{T_H} \simeq 8\pi M \left(1+\frac{\alpha^2}{20M^2}\right).\] 

Therefore, for fixed \(M\) and \(\omega\), the PDM scale suppresses the spectral emission rate: it reduces the effective emitting area in this approximation and increases the Boltzmann factor.

In the limit \(\alpha\to 0\), Eq.~\eqref{cc3} reduces to
\begin{equation}
\frac{d^2\mathbb{E}}{d\omega dt}=\frac{8 \pi^3 M^2 \omega^3}{\exp\left(8 \pi M \omega\right)-1}.\label{cc5}
\end{equation}

The second, and more appropriate, estimate in the high-energy regime is the shadow-based absorption cross-section \cite{Wei2013},
\begin{equation}
\sigma_{\rm lim}\approx \pi R_{\rm sh}^2,\label{cc6}
\end{equation}
where \(R_{\rm sh}\) is given by Eq.~\eqref{shadow_radius}. In the Schwarzschild limit, this gives
\begin{equation}
\left. \frac{d^2\mathbb{E}}{d\omega dt} \right|_{\rm Sch,\,sh} = \frac{54\pi^3 M^2\omega^3}{\exp(8\pi M\omega)-1},\label{cc7}
\end{equation}
because \(R_{\rm sh}=3\sqrt{3}M\). This distinction is important: Eq.~\eqref{cc5} corresponds to the horizon-area estimate, whereas Eq.~\eqref{cc7} corresponds to the asymptotic shadow/absorption estimate. The latter is more appropriate when discussing the emission rate in connection with the black-hole shadow.

The spectral emission rate is plotted in Fig.~\ref{fig:emission_rate}. Panel~(a) uses the near-horizon estimate \(\sigma_{\rm lim}\approx\pi r_h^2\), whereas panel~(b) uses the shadow-based high-energy estimate \(\sigma_{\rm lim}\approx\pi R_{\rm sh}^2\). In both cases, increasing \(\alpha/M\) suppresses the emission relative to the Schwarzschild case. The shadow-based curves are larger in magnitude because the high-energy absorption cross-section is controlled by the critical impact parameter rather than by the horizon radius alone.

\section{Conclusion}\label{sec:5}

In this work, we investigated the thermodynamic and radiative properties of a singularity-free black hole sourced by primordial dark matter and modeled effectively through a DBI scalar field. We used the metric normalization in which the integration constant \(M\) coincides with the ADM mass and the asymptotic expansion correctly reproduces the Schwarzschild term \(1-2M/r\). With this normalization, the ADM mass is related to the horizon radius by \(M=\alpha/[3B(r_h/\alpha)]\), and the Schwarzschild relation \(M\simeq r_h/2\) is recovered for \(\alpha/r_h\ll1\).

We derived the Hawking temperature and showed that the PDM scale lowers the temperature relative to the Schwarzschild value at fixed mass,
\[
    T_H\simeq
    T_S
    \left(1-\frac{\alpha^2}{20M^2}\right).
\]
For fixed \(\alpha\), the mass has a lower bound \(M_{\rm min}=2\alpha/(3\pi)\). In the limiting configuration \(r_h\to0\), the areal radius remains finite, \(\rho_h\to\alpha\), and the Hawking temperature tends to the finite value \(T_H\to1/(\pi^2\alpha)\).
We also discussed the entropy obtained from the fixed-\(\alpha\) first law. The result is not assumed to be exactly \(A/4\); rather, it is determined by an integral expression. In the large-radius regime, the leading term reproduces the Schwarzschild entropy, while subleading corrections depend on the regularity scale \(\alpha\).

The fixed-\(\alpha\) specific heat capacity was computed explicitly. We found that \(C_\alpha<0\) throughout the physical branch \(r_h/\alpha>0\). Hence, for the metric normalization and thermodynamic prescription adopted here, the regular PDM scale does not generate a locally stable canonical phase and no Davies-type divergence appears in the physical domain.

We also analyzed the sparsity of Hawking radiation. Using the effective-area prescription \(A_{\rm eff}=27A_{\rm BH}/4\), the geometrical sparsity parameter approaches the Schwarzschild value in the limit \(\alpha\to0\). In the perturbative regime \(\alpha\ll2M\), it receives the correction
\[
    \eta
    =
    \left(1-\frac{\alpha^2}{20M^2}\right)
    \eta_{\rm Sch}
    +
    \mathcal{O}\!\left(\frac{\alpha^4}{M^4}\right),
\]
which corresponds to a slight decrease in the intermittency of the Hawking flux within the adopted prescription.

Finally, we studied the spectral energy emission rate. The PDM scale suppresses the emission rate relative to the Schwarzschild case by reducing the effective horizon-scale emitting area and increasing the Boltzmann factor. We also emphasized the distinction between the near-horizon estimate \(\sigma_{\rm lim}\approx\pi r_h^2\) and the shadow-based high-energy estimate \(\sigma_{\rm lim}\approx\pi R_{\rm sh}^2\). The latter is the appropriate cross-section when the emission rate is interpreted in terms of the black-hole shadow.

The present analysis prepares the ground for a detailed numerical study of the temperature profile, heat capacity, sparsity parameter, photon-sphere radius, shadow radius, and emission spectra as functions of the PDM scale. These quantities will make the physical impact of the dark-matter regularization more transparent and will be especially useful for comparing this model with other regular black-hole geometries.

\scriptsize

\section*{Acknowledgments}

F.A. acknowledges the Inter University Centre for Astronomy and Astrophysics (IUCAA), Pune, India for granting visiting associateship. E. O. Silva acknowledges the support from Conselho Nacional de Desenvolvimento Cient\'ifico e Tecnol\'ogico (CNPq) (grant 306308/2022-3), Funda\c c\~ao de Amparo \`a Pesquisa e ao Desenvolvimento Cient\'ifico e Tecnol\'ogico do Maranh\~ao (FAPEMA) (grant UNIVERSAL-06395/22), and Coordena\c c\~ao de Aperfei\c coamento de Pessoal de N\'ivel Superior (CAPES) -- Brazil (Code 001).

%


\end{document}